# Cost Reducing Adiabatic Compressed Air Energy Storage for Long Duration Energy Storage Applications

Danlei Yang[1], Yang Wang[2], Jihong Wang[3], Zhenhua Rui[4,*], Wei He[1,*,**]

1   Department of Engineering, King's College London, United Kingdom
2 Energy System Catapult, United Kingdom
3 School of Engineering, University of Warwick, United Kingdom
4 College of Carbon Neutrality Future Technology, the China University of Petroleum (Beijing), China

*Corresponding author: ruizh@cup.edu.cn; wei.4.he@kcl.ac.uk;
**Lead Contact: Wei He

**Summary**
Long-duration energy storage (LDES) is vital for decarbonizing the energy system but faces economic challenges, including high upfront costs, low trading frequency, and limited revenue in current electricity markets. Compressed Air Energy Storage (CAES) is a promising LDES solution, though its economic viability, especially for long storage durations beyond lithium-ion battery capabilities, remains unclear. To address this, here we compiled and analyzed a global emerging adiabatic CAES cost database, showing a continuous cost reduction with an experience rate of 15% as capacities scaled from 10MW to 100MW. Our lifecycle discounted cash flow analysis suggests that adiabatic CAES could achieve economic viability for 10-100 hour storage durations, particularly with optimal geological siting to lower storage costs. This economically viable LDES option will enable large-scale grid balancing and support renewable integration over multi-day periods, making it a valuable asset for advancing deep decarbonization of energy systems.

**Introduction**
Long duration energy storage (LDES) is vital for grid stability, enabling the storage of renewable energy for periods ranging from days to years [1]. LDES technologies typically include pumped hydro (PHES), compressed air (CAES), flow batteries, and iron-air batteries. Unlike lithium-ion batteries, which have been widely deployed due to significant cost reductions and their ability to capitalize on short-term electrical markets like frequency markets [2], the economic viability of LDES remains uncertain. LDES technologies generally require large-scale energy storage capacities for power generation over an extended period (such as a full cycle over several weeks or even months), leading to substantial upfront investments. Consequently, LDES plants trade electricity less frequently than short-duration storage plants, resulting in lower load factors than typically daily cycled lithium-ion battery storage. For revenue generation, energy storage plants participate in electricity markets operating through structured mechanisms such as the day-ahead and intra-day markets. The day-ahead market enables participants to submit bids in advance to set prices and maintain grid stability [3], while the intra-day market allows for real-time adjustments to demand fluctuations [4,5]. Together, these mechanisms ensure reliable power supply and enhance market competitiveness. Compared to short-duration energy storage, which cycles frequently and participates in multiple electricity markets, LDES is traded infrequently and typically engages in fewer markets. Consequently, low revenue traded, lower load factors, and high initial investments often result in unclear economics, significantly limiting the widespread adoption of LDES technologies.

These typical operational characteristics and market revenue structures impose strict cost constraints on LDES technologies to be cost-effective. Among them, CAES is often considered one of the most economical options, with costs ranging from 28-295 $/kWh [6,7], largely due to different

designs with varied capacities between power and energy storage. CAES generates pressurized air using compressors, stores it in containers or underground caverns, and releases it to drive turbines that generate electricity. Two conventional diabatic CAES plants (320 MW Huntorf and 110 MW McIntosh), which combust a mixture of natural gas and compressed air before expansion, have operated for decades. Recently, adiabatic CAES (A-CAES) systems have been proposed to eliminate the need for natural gas by storing compression heat and reusing it during air expansion, thereby ensuring power capacity and improving energy efficiency.

CAES uses compressors and turbines for charging and discharging power, while typically relying on underground salt caverns to store large volumes of pressurized air. Salt caverns, which have been used for natural gas storage for decades, are not only compatible with compressed air but also suitable for storing other low-carbon gases such as hydrogen and carbon dioxide [8]. Previous research suggests that sufficient underground resources exist in many regions to support net-zero transitions [9]. However, despite its favorable costs and resource availability, CAES is frequently overlooked in energy transition studies involving storage technologies [10-12].

The learning or experience curve approach provides valuable insights into future cost estimates by visualizing and leveraging how investment efficiencies improve with increased production and deployment experience. This concept has been effectively applied in fields like photovoltaics [13] and lithium-ion batteries [14], where it has successfully tracked and predicted cost reductions for the past decades [11]. Projected cost estimates derived from learning curves are critical inputs for energy system models, enabling modelers such as those in the IEA [15] and policymakers [16] to analyze transition scenarios and optimize pathways to net zero [17].

However, the varying visibility of cost data across different energy storage technologies presents a significant and largely unaddressed challenge. While historical costs for technologies like lithium-ion batteries are well-documented across various applications—from Electronics, EV, residential to the utility [11] —such data is particularly scarce for LDES technologies, such as CAES. This unequal data availability and inconsistent assumptions behind future cost projections can lead to divergent energy transition pathways. Schmidt et al. [10] conducted a comprehensive assessment of the application-specific lifetime costs of storing electricity from 2015 to 2050 across various electrical markets and services. Their findings showed that in 2015, PHES and CAES dominated most applications, except those requiring quick response times and smaller sizes. However, by 2030, lithium-ion batteries are expected to become the most cost-efficient option for nearly all stationary applications. This shift is primarily due to the projected cost reduction rates: lithium-ion batteries are expected to decrease by 77% by 2030 and 86% by 2050. In contrast, CAES investment costs are assumed to remain unchanged by 2030 and increase by 2% by 2050. This stark contrast in projected cost trajectories highlights the critical importance of cost data in determining the cost-optimal solutions during the transition.

Is CAES a mature technology with limited potential for significant cost reductions? This study seeks to answer this question by addressing the data and insight gaps surrounding emerging A-CAES technologies. The lack of comprehensive data poses challenges for policymakers [18], who rely on highly credible information to make informed decisions. Without reliable cost estimates, the potential benefits and trade-offs of CAES for LDES applications remain uncertain, complicating the development of effective policies and incentives.

To address this gap, we systematically gather comprehensive cost data on A-CAES technologies and projects across the globe and analyzed the experience rate of A-CAES projects to date for forecasting possible future costs. This data is then used to analyze the cash flow of A-CAES projects across various storage durations, helping to identify economically viable services that are well-suited to CAES from now to the future. We explore strategies to accelerate cost reductions in CAES and discuss potential pathways for the cost-effective deployment of CAES systems for LDES applications, contributing to net-zero transitions.

**Results**
**Cost data of A-CAES projects and experience rate of the technology**
We search data points of CAES projects using reported capital costs or product prices and cumulative installed capacities based on data from peer-reviewed literature, research and industry reports, news items, energy storage databases and interviews with manufacturers (details in Table S1). Reported capital costs for the emerging A-CAES technology is highlighted, and all CAES projects are for grid-scale energy storage applications, as shown in **Error! Reference source not found.**.

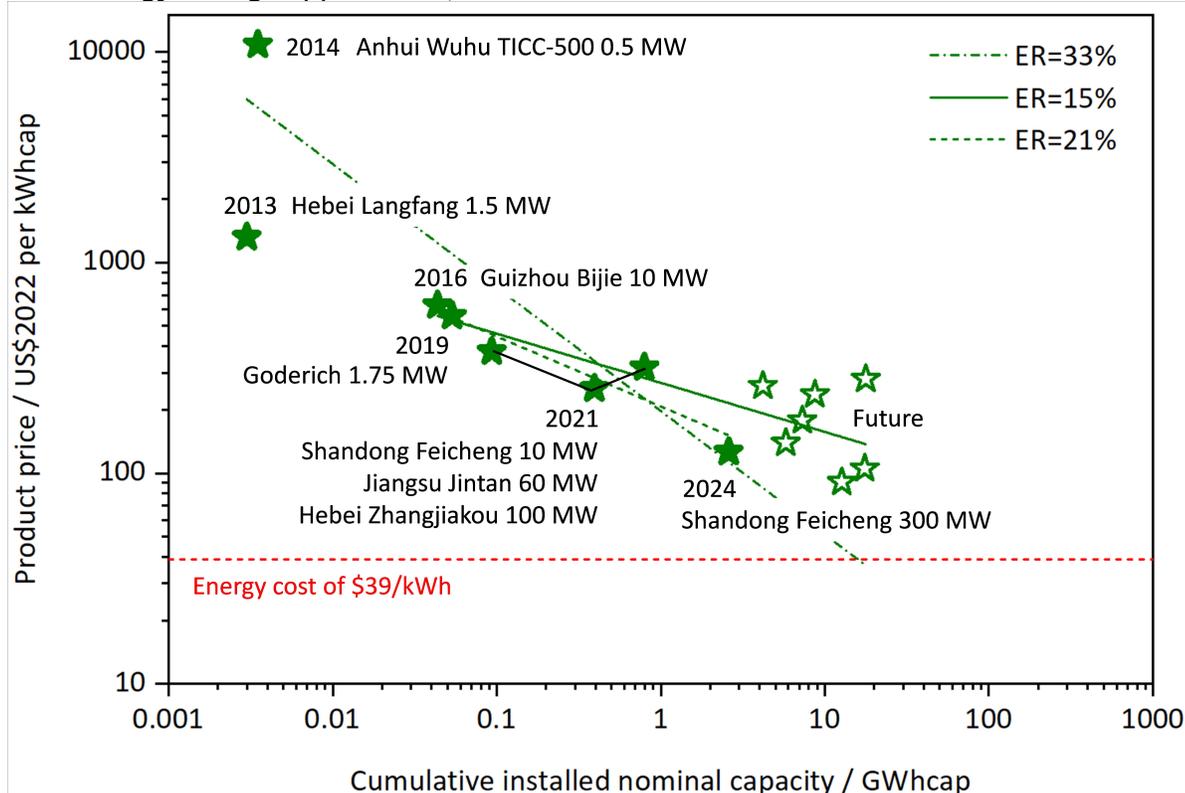

*Fig. 1 Cost data of A-CAES projects and associated experience rate analysis. This graph shows the decline in unit costs with costs in 2022 US dollars per megawatt-hour. The x-axis represents cumulative installed capacity in gigawatt-hours, and the y-axis shows product prices. Solid stars mark operational projects, while hollow stars indicate projects under construction or planned. Projects listed by year denote initiatives started that year. The green lines illustrate different learning curves: the dashed-dotted line represents all projects (2014 Anhui Wuhu TICC-500, 2021 Jiangsu Jintan, 2021 Hebei Zhangjiakou, and 2024 Shandong Feicheng, like constructed etc); the solid line excludes projects from 2013 and 2014; and the dashed line includes only completed (constructed) projects, also excluding those from 2013 and 2014. These curves collectively highlight the general trend of decreasing costs with increasing cumulative capacity. It is important to note that the 2013 project lacks specific capacity data. For analysis purposes, we assume a C-rate of 2 for this system. The exclusion of 2013 and 2014 projects in the solid and dashed green lines aims to eliminate the influence of this assumption, as well as the high costs associated with a 2014 laboratory-scale pilot project. The red dashed line marks the $39/kWh energy cost from [10], which is the investment energy cost for compressed air energy storage in 2015.*

A detailed list of all found CAES projects can be found in Table S2. A-CAES was reported to be as low as $120/kWh for a 100+ MW project in 2024 and exhibits clear a cost reduction trend with possible

market average prices below US$157/kWh above 10 GWh, and US$92/kWh above 100 GWh, installed.

A validated learning curve model (validated with existing datasets of other technologies in the literature, as shown in Table S3) is used to estimate the experience rate (ER) of A-CAES. The analysis of A-CAES project cost data produced two ERs: a 33% ER when including all project costs and a 15% ER when considering only larger-scale projects (above 10 MW). To minimise the influence of extreme values and scale-related performance variations, sub-MW and early experimental A-CAES projects were excluded from the dataset, retaining only ≥10 MW grid-scale cases. High costs from small-scale, experimental projects in the early 2010s likely inflated the overall ER, making the 15% rate more reflective of anticipated future deployments (tens to hundreds of MW), and thus it is used in this study.

Looking into the cost trend, CAES project costs have declined from over $10,000 per kWh in 2013 to about $120 per kWh by 2024, largely due to increased project scale and technology maturity. Early projects with small capacities (0.5–1.5 MW) were costly due to experimental status, but larger-scale projects (10–100 MW) now achieve much lower costs (~$200/kWh). As additional high-capacity projects (200-500 MW) are planned, costs are expected to decrease further to close to $100/kWh.

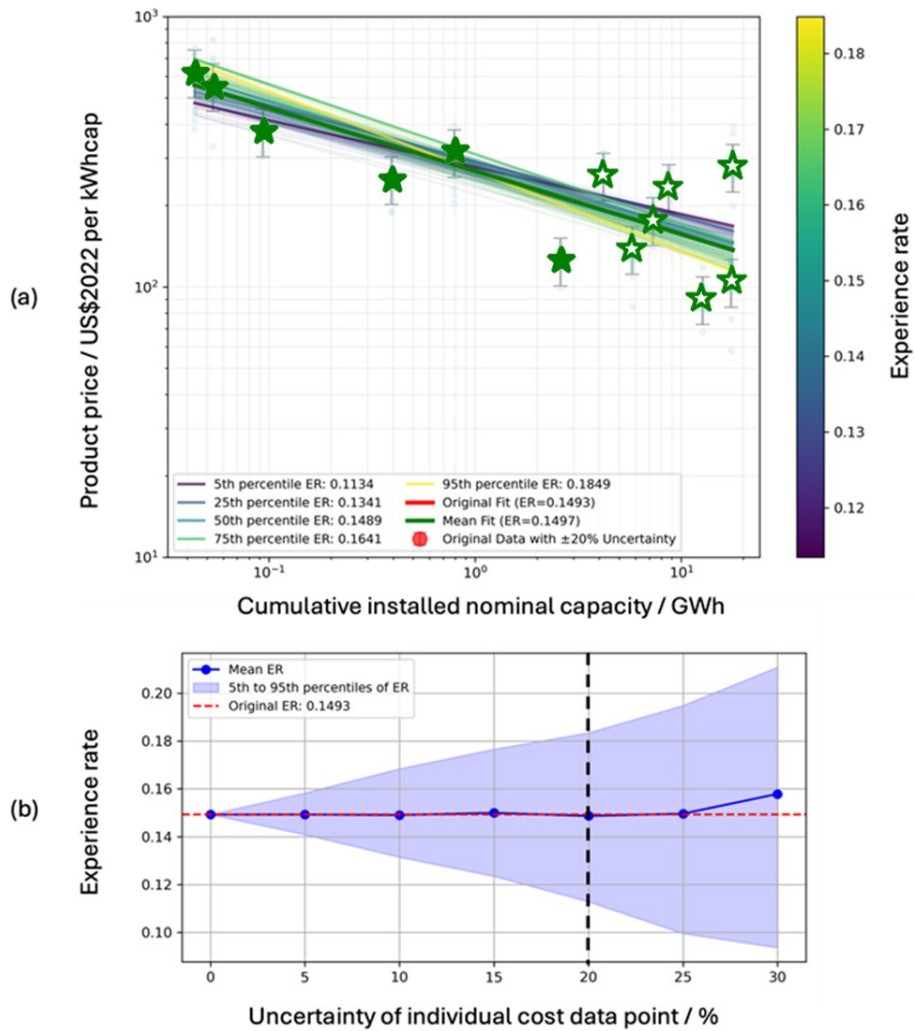

Fig. 2 Uncertainty analysis of the cost data. (a) Wright's Law curve fits under ±20% data uncertainty.

*The red circles show the original cost data, each with vertical error bars representing ±20% uncertainty. The multitude of green to yellow lines indicate different Monte Carlo realizations of Wright's Law, where each realization randomly perturbs the original data within the specified ±20% range before re-fitting. The color scale on the right corresponds to the fitted experience rate. Several highlighted curves (labeled 5th, 25th, 50th, 75th, and 95th percentile) show how the central 90% of ER values span roughly 11%–18%. The thick red line is the original best-fit curve, with its ER of 15% indicated in the legend. (b) Sensitivity analysis of the experience rate as the assumed uncertainty in each cost data point increases from 0% to 30%. The horizontal axis represents the percentage uncertainty assigned to each cost observation, while the vertical axis shows the resulting ER distributions from repeated Monte Carlo runs. The blue dots indicate the mean ER at each uncertainty level, and the shaded region spans the central 90% interval (5th–95th percentile). The dashed vertical line at 20% aligns with the assumptions used in panel (a).*

To analyze impact of embedded uncertainty in the collected datasets, we adopt a random sampling approach through Monte Carlo simulation (see Method), applying a ±20% variation to each data point. This ±20% assumption serves as a baseline estimate to reflect the general variability in cost reporting across both constructed and planned projects. For constructed projects, this range accounts for incomplete public data, inconsistent reporting scopes, and the lack of third-party validation. For instance, the reported cost of the Goderich plant by Hydrostor varies by a factor of three (from $1000 to $3000/kW), due to the absence of detailed cost breakdowns or independent verification. This process generates multiple noisy realizations of the dataset, each of which is refitted to Wright's Law to obtain a distribution of ER. Fig. 2(a) shows that, within the central 90% interval (from the 5th to the 95th percentile), the fitted ER range from roughly 11% to 18%, illustrating how parameter estimates can shift under plausible levels of data uncertainty.

For planned projects, we acknowledge that ±20% may be a conservative estimate, as such projects often experience more volatile and uncertain cost trajectories. Therefore, we further examine uncertainty levels from 0% to 30% and find that, as shown in Fig. 2(b), the spread in the ER distribution increases with larger assumed data-point uncertainty. At ±30%, the 5th to 95th percentile range for ER is about 10% to 20%. Even though we do not have direct measurements of individual data-point uncertainty, exploring a reasonably wide range of possible errors via Monte Carlo gives us confidence about the robustness of these findings. In all cases, the resulting ER remains consistently positive and generally exceeds 10%, implying that, despite unknown measurement accuracy, a two-digit ER is plausible, and the underlying cost-reduction trend holds.

To verify the reliability of the cost data from another angle, we also examine bottom-up cost modelling from the literature. These studies show consistent estimates for pilot- and full-scale A-CAES projects, aligning well with the claims made by industry and early stage demonstrators. For instance, a 300 MW A-CAES plant is reported to have a total investment cost of $245.11 million, while larger capacities (1,500–5,000 MWh) require $245.26–$573.86 million ($163.51/kWh- 114.77/kWh) [19]. For a 10 MW / 80 MWh AA-CAES system, the static construction cost is approximately ¥66.79 million. Further calculations indicate that for storage capacities between 160 MWh and 800 MWh, the initial investment ranges from ¥130.37 million to ¥607.86 million ($121.61/kWh-113.41/kWh in 2022 USD) [20]. The unit cost of an A-CAES system was estimated at $334/kWh in 2013 ($417.17/kWh in 2022 USD) [21]. In a preliminary capital cost assessment for Norway, the estimated investment cost is 2,700 NOK/kWh ($281.25/kWh in 2022 USD) [22]. These economic models estimate the cost of A-CAES to range from around $100/kWh to $400/kWh, aligning with the actual project cost range obtained in Fig. 1.

Therefore, our analysis underscores clear cost reductions in A-CAES development, with an estimated ER around 15%. We note that Schmidt et al. [10] did not empirically fit an ER for CAES, but instead assumed that CAES would follow the ER of pumped hydro storage (PHS) reported in their earlier work, which had high uncertainty. This assumption may not be appropriate for emerging fully electrified A-CAES technologies, which differ substantially from PHS in maturity and cost drivers. Schmidt et al.'s baseline CAES costs remain lower than our findings, suggesting that despite recent cost declines, A-CAES is still more expensive than previously assumed. This discrepancy is due to a lack of comprehensive cost data specific to A-CAES projects and confusion between conventional natural gas-integrated D-CAES and emerging fully electrified A-CAES systems. Earlier analyses often combined costs from traditional D-CAES projects in Germany and the USA with those of newer A-CAES technologies, leading to inconsistencies in both cost and technology maturity assessments. Although A-CAES is less mature than some prior studies assumed, it has made significant progress over the past decade, with increased scalability and cost reductions positioning it as a promising option for long-duration applications.

Additionally, A-CAES has been demonstrated at scales from kilowatts to megawatts [18], with the TICC-500 unit validating the feasibility of MW-scale operation and providing key performance data [23]. Subsequently, system capacities have expanded to tens or even hundreds of megawatts, with storage durations of up to 10–15 hours [24]. Unlike modular technologies such as lithium-ion cells or photovoltaic modules, whose performance remains constant across scales, A-CAES shows improved efficiency and operational stability as capacity increases [25], driven by thermodynamic optimisation, better component matching, and proportionally reduced thermal losses. Large-scale plants can deploy compressors, expanders, and thermal energy storage (TES) systems optimised for specific temperature and pressure ranges, further enhancing performance [26]. Modelling studies indicate that small-scale prototypes typically achieve round-trip efficiencies (RTE) of 40–55%, whereas optimised large-scale designs can exceed 70% [24, 26], as demonstrated by a 35 MW system achieving ~70% RTE with 15 hours of continuous operation [24]. This scale-driven performance improvement not only influences cost structures but also accelerates the rate of unit cost reduction with cumulative installed capacity, underscoring the need to account for scale effects in learning curve analysis.

**Learning curve of CAES in comparison to other electrical energy storage technologies**
In comparing the cost curves of CAES technology, including both A-CAES and D-CAES, with all other energy storage technologies, lead to insights of factors that further drive the cost reduction, as illustrated in Fig. 3.

The figure highlights two D-CAES projects, Huntorf and McIntosh, both with low costs similar to estimates by Schmidt et al. [10]. McIntosh, with a large storage capacity of 2,640 MWh, shows particularly low unit costs. This demonstrates the benefit of CAES systems, where energy and power storage are decoupled, allowing for large-scale energy storage at reduced specific costs due to the dominance of air storage costs—a feature also relevant to A-CAES projects. However, with limited data points, calculating an ER for D-CAES projects lacks statistical significance.

A-CAES, with an ER of around 15%, is comparable to lithium-ion batteries (12%), electrolyzers (19%), and fuel cells (21%). Comparing with these energy storage technologies, CAES start to benefit from some recognized cost reduction drivers demonstrated by lithium-ion batteries, electrolysis, and fuel cells, such as R&D, learning-by-doing, and economies of scale. However, the mechanisms through which these reductions are realized differ due to each technology's unique characteristics. For example, lithium-ion batteries achieve cost declines through advancements in cell chemistry and materials science, leading to higher charge densities and lower material costs [14]. Electrolysis benefits from manufacturing improvements and innovations like solid oxide electrolyser cells and zero-gap alkaline electrolysers [27, 28]. Fuel cells reduce costs via accumulated learning and scaling effects [29]. These pathways highlight that while the drivers are similar, the specific areas of focus for cost reduction are distinct. In 2022, the U.S. Department of Energy (DOE) estimated storage costs at

$143–263/kWh for pumped hydro, $356–405/kWh for lithium-ion, $385/kWh for vanadium flow, and $409/kWh for lead-acid batteries. With global lithium-ion battery capacity reaching 170–340 GWh by 2023 (IEA), experience curve effects suggest costs could decline to $300–400/kWh, highlighting the greater relevance of ER-based analysis over static cost figures.

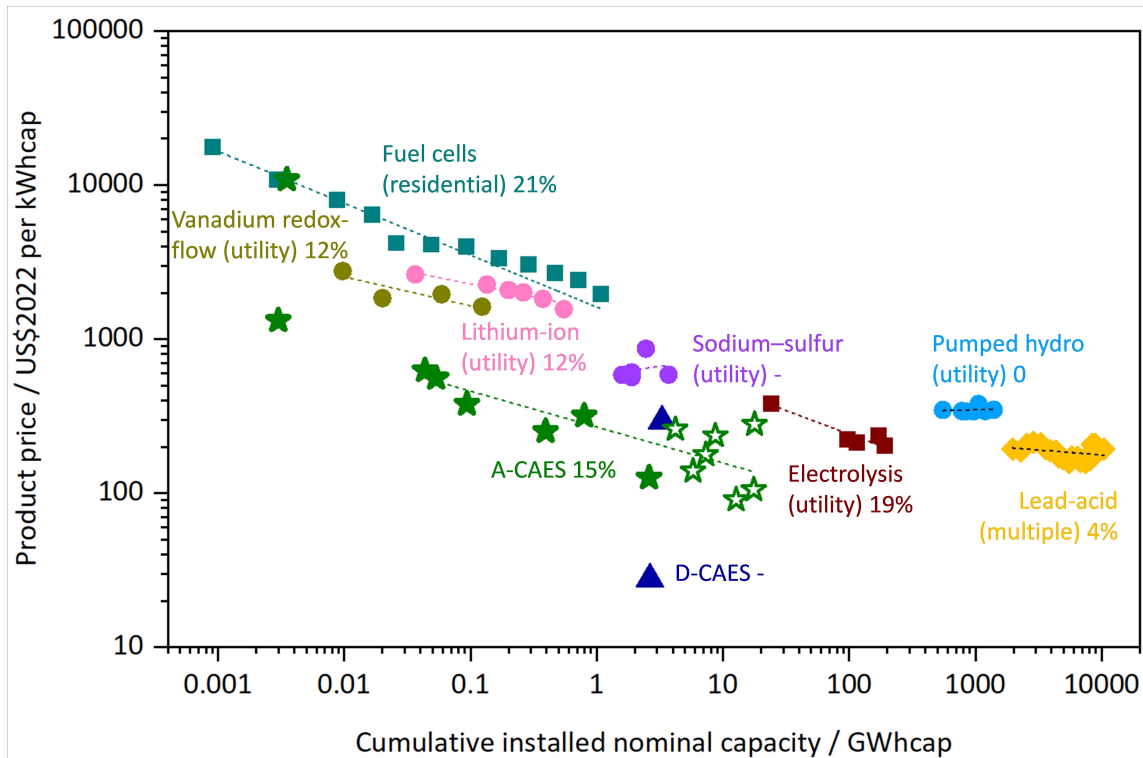

Fig. 3 Learning curve of CAES and comparison with other energy storage technologies. This graph compares the product price of various energy storage technologies in 2022 US dollars per kWh capacity, plotted against cumulative installed nominal capacity in GWh. The technologies include pumped hydro, lead-acid, lithium-ion, sodium-sulfur, vanadium redox-flow, electrolysis, fuel cells, D-CAES (diabatic CAES), and A-CAES (adiabatic CAES). The x-axis represents cumulative installed capacity, while the y-axis shows product price. Data of non-CAES technologies is from [11]. The graph highlights a general trend of decreasing costs as installed capacity increases, with distinct cost curves for each technology. Dotted lines represent the learning curves for different storage technologies with their ER.

Historical evidence indicates that other large-scale, non-modular technologies have also exhibited learning-curve effects in their early deployment phases [30]. For instance, pumped hydro storage (PHS) experienced only modest early cost declines before reaching technological maturity, after which capital costs stabilised and learning rates became less predictive. In contrast, nuclear power in countries such as France and the United States achieved substantial early cost reductions through standardised reactor designs and strong "learning-by-doing" effects, enabling learning rates to successfully predict future costs during expansion.

Transferable learning factors from these technologies to A-CAES include economies of scale and manufacturing improvements. Just as scaling up production has reduced costs for batteries and electrolysers, increased deployment of A-CAES could lower unit costs of compressors, turbines, and other components. Technological innovations in materials and component designs, supported by continues R&D investments, can also enhance A-CAES efficiency and reduce costs, mirroring advancements in other energy storage technologies.

CAES has already achieved notable cost reductions by leveraging economies of scale and manufacturing improvements, particularly evident in China, where established manufacturing and supply chains for large-scale A-CAES projects have accelerated growth. In 2024 alone, over 11 projects that have over 100 MW capacity have been announced (a total capacity of 13.95 GWh) [31], showcasing rapid expansion. Scaling up production, similar to the cost reductions achieved in batteries and electrolysers, is driving down the unit costs of compressors, turbines, and other essential components for CAES. Additionally, increased competition among technology suppliers further stimulates R&D, resulting in innovations across A-CAES designs, manufacture, and supply chains.

There is a clear trend toward higher operating temperatures in A-CAES projects, which aligns with thermodynamic principles. According to the second law of thermodynamics, higher temperature differentials improve thermal-to-electric conversion efficiency. As projects scale from 10 MW to over 100 MW, thermal storage systems have evolved from pressurized water tanks, used in early TICC-500 project in 2014 [23], thermal oil at a temperature of around 300°C used in the 60 MW project in Jiangsu China in 2021 [32], to molten-salt-based storage operating around 500°C in the 350 MW project in Shandong China started to build in 2023 [33]. Note that, Molten-salt-based TES at ~500 °C can experience significant static thermal losses, which, without proper design, may limit the economic viability of multi-day or seasonal LDES. These losses can be mitigated by optimizing tank geometry and insulation. Potential high-temperature TES pathways include composite phase change materials (CPCM) -enhanced molten salts [34], ceramic or brick sensible heat storage operating at ~1000 °C [35], and metal-based latent heat storage above 1000 °C [36]. Additionally, high-temperature CAES designs in the literature indicate the potential for using hydrogen as the fuel of the combustion chamber instead of gas, further increasing operating temperatures and efficiency [37]. These advancements likely lead to continuous system capacity scale up, and likely cost reduction per unit energy and power delivered.

A-CAES faces unique challenges that distinguish it from battery and electrochemical energy technologies. Its reliance on suitable geological formations for large-scale air storage introduces cost variability and limits site selection, unlike the modular and location-independent nature of batteries and fuel cells. Additionally, compressors and turbomachinery are mature technologies in conventional applications such as gas turbines, with established supply chains. However, these designs are not fully optimized for A-CAES applications, where part-load flexibility, high-pressure operation, and integration with thermal storage introduce design challenges. Therefore, while conventional turbomachinery has limited room for cost reduction, A-CAES-specific designs still offer significant opportunities for further innovation and cost optimization. To fully realize its cost-reduction potential, CAES must address these unique factors by optimizing storage methods and aligning deployment with existing supply chains.

To overcome geological constraints, R&D is needed to explore alternative storage solutions, such as low-cost above-ground tanks or engineered subsurface storage, to make air storage more adaptable and affordable. Early attempts, like the US venture Lightsail's carbon fibre tanks, underscored the need for cost-effective, geology-independent storage, despite limited success commercially [38]. Excavated rock caverns, while more costly than salt caverns, are also gaining interest due to their flexibility in expanding storage options [39].

Although the designs of compressors and turbomachinery for the CAES technology remain underdeveloped, integrating tailored designs into current manufacturing capabilities presents an opportunity for near-term cost reductions. However, substantial long-term savings will likely require further innovation in component design. By addressing these areas—optimizing air storage, developing containment structures, and streamlining deployment processes—A-CAES can strengthen its competitiveness in the energy storage market.

**Economics and Operation of A-CAES for Long-Duration Energy Storage Applications**
In this section, we explore the potential of A-CAES to provide extended duration energy storage and examine its associated power. A lifecycle discounted cash flow analysis [1] is used to assess the financial viability and cost-reduction potential of CAES systems across different storage durations. By comparing the present value of revenues (from typical energy and capacity services) with the total cost of ownership (including capital, operational, and maintenance costs) over the project's lifecycle, the cash flow analysis approach determines whether CAES projects are economically viable. If revenues exceed costs, the project is viable; if not, it is unviable economcially. This lifecycle discounted cash flow analysis offers insights into the economic viability of CAES, helping to identify the most suitable applications, especially those requiring long storage durations.

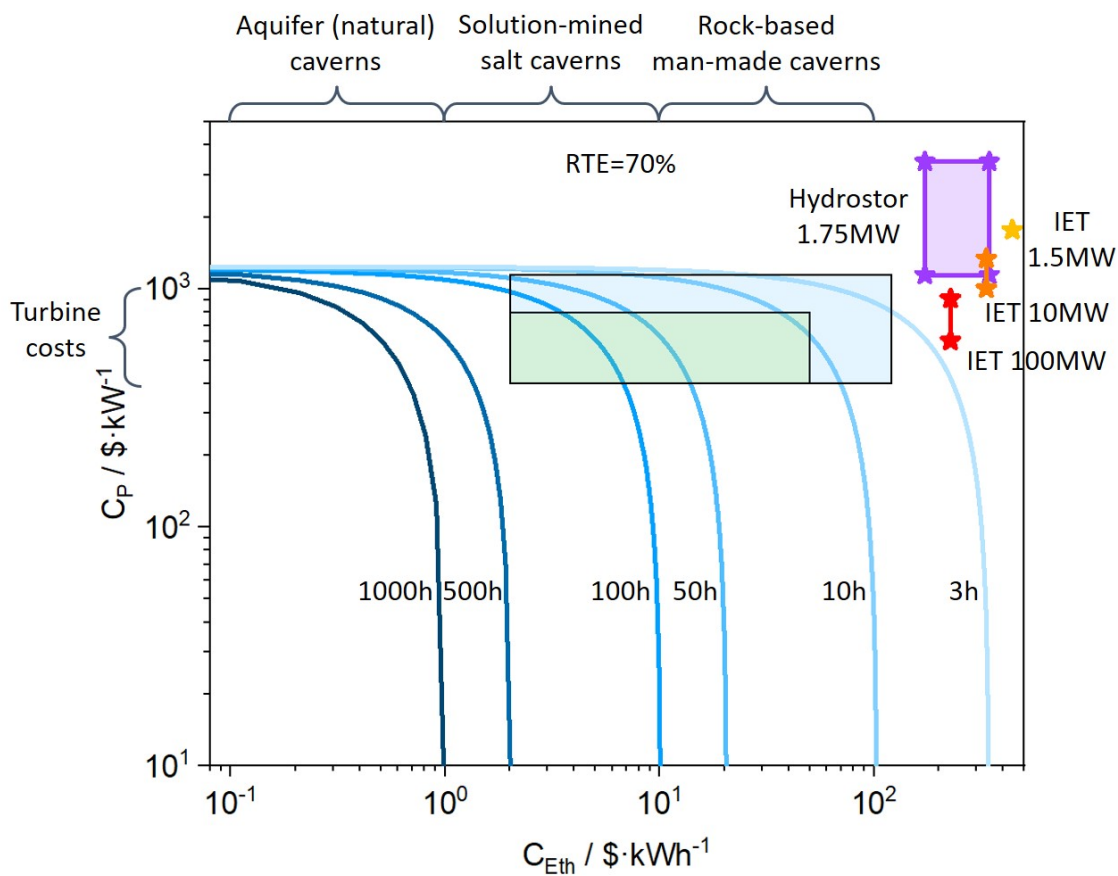

*Fig. 4 Economics of A-CAES for long-duration energy storage: Power and energy cost trends by discharge duration. This chart shows the relationship between energy cost ($C_{Eth}$ in \$/kWh) on the horizontal axis and power cost ($C_p$ in \$/kW) on the vertical axis for various energy storage technologies, focusing on Compressed Air Energy Storage (CAES). The plot provides insight into how different discharge durations impact the cost structure of CAES systems. First, the series of blue curves represent various CAES systems with different discharge durations, ranging from 3 hours to 1000 hours. The darker curves correspond to longer durations (e.g., 500h and 1000h), while lighter curves correspond to shorter durations (e.g., 3h, 10h, and 50h). As the duration increases, the energy cost ($C_{Eth}$) decreases, but the power cost ($C_p$) increases. This trade-off reflects the capital-intensive nature of CAES installations with longer discharge durations. Second, the analysis assumes a round-trip efficiency (RTE) of 70%, which is consistent with*

*values commonly adopted in peer-reviewed modelling studies and represents a plausible upper bound for optimized A-CAES systems [12, 18, 24, 49]. The lightest blue curve represents a 3-hour discharge duration, which aligns closely with real-world CAES projects, as indicated by the Hydrostor [25] (purple box) and IET [41] projects (red/yellow/orange stars). Third, Hydrostor (1.75 MW, marked in purple) represents a commercial CAES system, showing its actual power and energy costs. IET projects (marked in red, orange, and yellow stars) indicate different CAES installations with varying power and energy costs. These markers help validate the modelling results by comparing them with actual deployment data. Fourth, the discounted cash flow (DCF) model used for these cost estimations is explained in Supplementary Information Table S4. Further validation of the DCF methodology is provided in Table S5, ensuring that the modelled cost trends align with economic expectations. Fifth, the graph effectively shows that shorter discharge durations (e.g., 3h-10h) tend to have higher energy costs but lower power costs, making them more competitive for applications requiring frequent cycling. Conversely, longer-duration CAES systems (e.g., 500h-1000h) have significantly lower energy costs but higher power costs, indicating their potential for seasonal storage rather than daily cycling. The plotted real-world projects (Hydrostor and IET) fall within the expected cost range, confirming that the model captures practical cost structures of existing CAES systems. Figure S1 and Figure S2 illustrate the trends of power cost and energy cost for CAES as a function of discharge duration, evaluated at round-trip efficiencies of 60% and 70%.*

Fig. 4 illustrates the power cost ($C_P$) and energy cost ($C_{Eth}$) of A-CAES technologies across various storage durations (e.g., 3, 10, 50, 100, 500, 1000 hours) and round-trip efficiencie (RTE) of 70% that represents the latest A-CAES development. Each duration curve demonstrates the relationship between cost, storage duration, and efficiency, providing insights into the economic viability of different A-CAES configurations. The claimed costs of IET's projects (1.5 MW, 10 MW, and 100 MW capacities) and Hydrostor's 1.75 MW project are plotted. The placement of these projects along the duration and RTE curves suggests near economic viability for A-CAES systems, particularly for applications with a 3-hour duration at their claimed RTE (around 70%), the current standard for IET and Hydrostor systems. This positioning indicates that A-CAES could potentially expand to longer-duration applications if further cost reductions are achieved.

For A-CAES projects to achieve economic viability at longer durations, energy costs, rather than power costs, will be the critical factor. Current A-CAES turbine power costs range from $600–900/kW, close to the commonly observed range of $400–500/kW based on our survey of major turbomachinery manufacturers, such as General Electric (LM5000 STIG at $410/kW installed) and Allison Engine ($562/kW installed) [40]. This cost premium reflects the lack of mature, commercially optimized turbomachinery tailored for A-CAES systems. As such, while general turbomachinery is mature, the specific application in A-CAES remains underdeveloped and offers potential for cost reduction, especially given the lower inlet temperatures and non-combustion environment typical in A-CAES designs. This margin leaves some potential for further cost reductions to significantly impact longer-duration applications. The cost curves suggest that reducing power costs alone would mainly benefit applications with a bit longer than 3h but still short-duration applications, typically up to 3–5 hours.

In contrast, energy storage costs have substantial potential for reduction, enabling economically viable applications with durations of over 10 hours and even up to 100 hours. Current energy costs for A-CAES projects are around $200–400/kWh, with IET's projects using above-ground stainless steel high-pressure tanks [41] and Hydrostor employing man-made underground rock-based caverns [42]. While these approaches aim to reduce geological constraints, they are costlier than conventional gas storage methods.

Air storage costs vary significantly depending on geological conditions. On the order of, natural porous rock formations can be as low as <$1/kWh, while solution-mined salt caverns can be

<$10/kWh [43]. Dry-mined salt caverns and abandoned rock mines are higher, and man-made rock caverns excavated from hard rock are even more expensive, ranging from $10–100/kWh [43]. Each option has trade-offs: natural aquifers lack proven gas-tightness and operaitonal safety at scale [44], salt caverns are mature but geographically distruted unevenly [45], and rock-based caverns are more widely available but more costly.

Reducing energy storage costs through optimal cavern selection and technological innovations is essential for A-CAES to acclerate the deployment of long-duration applications economically. Salt deposits, the geological resource for creating salt caverns, are abundant in many areas, including Western and Central Europe and parts of the USA [45], where renewable energy generation is rapidly expanding. Utilizing these resources could significantly support low-cost, LDES with salt-cavern-based A-CAES technologies. Alternatively, cost reduction in man-made lined rock caverns through further R&D will be key to expanding the applicability of A-CAES in the energy storage market [46].

**Discussions**
Our analysis indicates a 15% ER for A-CAES projects over the past decade, with costs as low as $120/kWh achieved for 100 MW+ deployments in 2024, driven by increased project scale and technology maturation. Early projects with small capacities (0.5–1.5 MW) were costly due to their experimental nature, but larger projects (10–100 MW) have reached much lower costs around $200/kWh. With additional high-capacity projects (200–500 MW) in the pipeline, further cost reductions are anticipated. Following the estimated 15% ER trend, market average prices could fall below $157/kWh for capacities above 10 GWh and $92/kWh for capacities above 100 GWh.

To realize this potential by continuously driving down the cost, our findings indicate both immediate and long-term cost reduction opportunities for A-CAES, enhancing its viability for medium-duration (10+ hours) to long-duration (100+ hours) storage applications. In the short term, standardizing components like compressors and turbines to improve the integration of current manufacture and supply chains, optimizing site selection for reducing energy storage costs, particularly using geological salt deposits for salt caverns can lower costs. These improvements would streamline manufacturing, reduce project timelines, and lower both capital and operational expenses, unlocking A-CAES economically viable to provide over 100-hour energy storage, being competitive in the current energy market.

Long-term continuous cost reductions will likely come from expanding low-cost storage options, particularly through innovations in cavern engineering including demonstration of natural geological resources and advancement of low-cost engineered rock-based caverns, would enable even longer duration beyond 100 hours applications and broader geographic deployment of A-CAES. Additionally, integrating A-CAES with renewable energy systems for grid stabilization services and leveraging economies of scale as more projects are deployed can drive down costs further. Policy support and financial incentives could accelerate these developments, fostering a more robust market for A-CAES and positioning it as a key technology for the energy transition.

Achieving 10-100 hour storage applications with A-CAES technology would be a transformative step for the energy system, enabling reliable LDES crucial for grid stability. Defined by organizations like the U.S. Department of Energy as storage beyond 8-24 hours [47], LDES fills a critical gap by balancing intermittent renewable sources to mitigate power outages for example due to unexpected weather event over an extended period. CAES is particularly well-suited for this role due to its low-frequency cycles and potential for long discharge durations, making it ideal for applications like seasonal energy storage. Our cash flow analysis suggests that A-CAES could economically support storage durations up to 500-1000 hours, if the storage cost can reduce to ~$1/kWh. By economically enabling these longer durations, A-CAES could significantly enhance grid resilience, facilitate a higher penetration of

renewables, and reduce reliance on fossil-fuel-based backup, advancing the transition to a more sustainable energy system.

The results presented in this study, is primarily subject to uncertainties in A-CAES cost data, which are drawn from various public sources (literature, reports, websites) with variable accuracy that are difficult to measure. Also, the lack of detailed information in some CAES projects, so assumptions were necessary to conduct the analysis. For instance, midpoint values were used if a range of cost was claimed by the project, and C-rates were assumed to estimate the storage duration of several projects if they are not explicitly defined (details are in Method). Additionally, while the learning curve approach provides useful insights, it relies on empirical correlations and affected by the quality and quantity of project data. Despite these limitations, the learning curve model is validated using data of more established technologies, and as A-CAES technology advances, more precise parameter settings and scenario modeling can refine and strengthen our conclusions.

**Limitations of study**
However, one major barrier to A-CAES deployment is the lack of transparent and consistent performance and cost data. As shown in Table S1, most A-CAES projects—especially pilot or early-stage systems—do not disclose key metrics such as round-trip efficiency, cycling behavior, or cost breakdowns. This data scarcity is due to several factors: many projects are still under development or operated privately without disclosure obligations; commercial confidentiality limits public reporting; and unlike mature technologies, A-CAES lacks standardized protocols for performance evaluation and reporting. In addition, A-CAES systems differ fundamentally from modular battery-based systems: while batteries are built from identical, scalable modules with well-defined metrics, A-CAES plants are typically bespoke, large-scale installations engineered case-by-case—more akin to thermal power plants than to modular electrochemical devices. As a result, detailed stand-alone component data is rarely reported. Similar data transparency challenges have been observed in other large-scale energy technologies. For example, Liu et al. (Nature, 2025) [48] adopted a comparable approach in collecting and analyzing nuclear power plant cost data, using a wide range of publicly available sources—including peer-reviewed literature, industry reports, environm[49]ental impact assessments, project documents, company prospectuses, and news reports—and noted that previous global nuclear cost datasets often relied on grey literature such as industry reports and news articles. To address this, greater transparency, third-party validation, and standardized data sharing will be essential for building confidence in the technology and accelerating adoption.

In summary, this study demonstrates A-CAES's strong potential as a cost-effective solution for LDES, with an estimated ER of 15% signaling continuous cost reductions. Our lifecycle discounted cash flow analysis suggests that A-CAES could achieve economic viability for storage durations of 10–100 hours with optimal geological siting. Furthermore, it could support durations of 100+ hours if storage costs drop to the order of $1/kWh, making it an ideal option for balancing renewable energy and enhancing seasonal grid resilience. As A-CAES technology advances, further data refinement will improve model accuracy, positioning A-CAES as a pivotal technology for a stable, low-carbon energy system.

**Acknowledgement**
We would like to acknowledge the support of UKRI-EPSRC [grant number EP/W027372/1]. W.H. would like to acknowledge the support of the Royal Academy of Engineering (RAEng) Engineering for Development Research Fellowship [grant number RF\201819\18\89].

**Declaration of Interests**
The authors declare no competing interests.

**Method details**

**Experience curves for electrical energy storage technologies**

This study utilizes experience curve analysis to project the future costs of compressed air energy storage (CAES) technology. This approach leverages historical data on product prices and cumulative installed capacities, sourced from academic literature, industry reports, and energy storage databases.

The study employs Wright's Law, which correlates cost reductions with cumulative production increases. Experience curves are developed to illustrate how prices decrease as production scales up [11].

$P(x) = A \times X^{-b}$

$ER = 1 - 2^{-b}$

Here, $P(x)$ represents the price per unit capacity, $X$ denotes cumulative installed capacity, $A$ is a normalization factor, $b$ is the experience rate parameter and $ER$ is the experience rate.

A linear regression of the logarithms of price and capacity data provides estimates for the experience rate, showing expected price reductions as CAES deployment increases.

Currency conversions are conducted in two steps. First, historical prices in local currencies are adjusted for inflation using OECD Consumer Price Indices. Next, these adjusted prices are converted to 2022 US dollars using OECD Purchasing Power Parity indices.

To convert data from power-based units (US$/kW, GW) to energy-based units (US$/kWh, GWh), the power-to-energy ratios (C-rates) specific to each technology are applied. The power-to-energy ratio defines the balance between the power output and energy storage capacity.

**Monte Carlo Simulation and Data Uncertainty Analysis**

Monte Carlo simulation is used to capture and propagate uncertainty. In the context of modeling cost data with Wright's Law, each measured data point is subject to observational and measurement errors. By applying Monte Carlo simulation here, we randomly sample noise (e.g., within a prescribed standard deviation like ±20%) and add it to the original data points many times, each time re-fitting the cost model. This procedure mimics the idea of repeatedly observing "slightly different" versions of the same underlying dataset, thereby creating a distribution of possible parameter fits. The spread and shape of that distribution reveal the model's sensitivity to uncertainty in the data.

Such repeated simulations provide tangible benefits for analyzing data reliability. First, if the resulting fitted parameters (like the scale factor A or the learning exponent b) vary only modestly across thousands of Monte Carlo runs, that means the original data—and the model's conclusion—are relatively robust to data's uncertainty. On the other hand, if the parameter estimates swing widely across the simulations, it signals a higher sensitivity, meaning that small data uncertainties can significantly shift the Wright's Law curve. This indicates caution when interpreting the results; the data may not tightly constrain the model, or the assumed functional form might not capture all the variability.

Moreover, by examining percentiles of the resulting parameter distributions or experience-rate ER values, one can gain insight into confidence intervals and identify "worst-case" or "best-case" scenarios. This in turn informs decision-makers about the level of risk or confidence in forecasts derived from the learning curve. For example, if the 5th and 95th percentile estimates for ER remain close, it suggests a narrow band of likely future costs, whereas a wide interval suggests more caution in predicting outcomes.

Ultimately, Monte Carlo simulation enriches our understanding of the model's reliability by exposing the range of plausible fits given the inherent uncertainty in the selected datasets, supported by reliable data analysis. While the baseline fit still offers a central estimate, seeing how that estimate changes under many randomized perturbations paints a more complete picture of both the strengths and potential fragilities of the data and model assumptions.

**Discounted Cash Flow Analysis Approach**

This study employs discounted cash flow analysis to evaluate the financial viability and cost reduction potential of compressed air energy storage (CAES) systems. The analysis uses historical data and various economic models to project future costs and revenues.

The discounted cash flow (DCF) analysis is used to determine the financial feasibility of CAES by equating the present value of revenue (PVR) with the total cost of ownership (TCO) over the system's lifecycle. The fundamental equation is:

PVR = TCO

The present value of revenue (PVR) is calculated by considering the different revenue streams generated by CAES systems, such as arbitrage and capacity market remuneration. The total cost of ownership (TCO) includes all costs associated with the CAES system, such as capital costs, efficiency losses, and operating expenses.

To provide a comprehensive view of cost effectiveness, the energy-normalized DCF equation considers both energy and power contributions [1]:

$$\sum_{t=1}^{T}(1+r)^{-t}[\Delta_{E,t}n_{c,t} + R_{P,t}d^{-1}] =$$

$$C_P d^{-1} + C_{E,th}\eta_D^{-1} + \sum_{t=1}^{T}(1+r)^{-t}[n_{c,t}P_{C,t}(\eta_{RTE}^{-1} - 1) + n_{c,t}VOM_t + d^{-1}FOM_t] + C_{Re}(1+r)^{-L/n_c}$$

where, $\sum_{t=1}^{T}(1+r)^{-t}$ is summation over the project term $T$, discounting each term by the discount rate $r$, $\Delta_{E,t}$ is frequency and average price differential of charging and discharging modes (indicative of arbitrage applications), $n_{c,t}$ is number of cycles in year $t$, $R_{P,t}$ is revenue from capacity payments in year $t$, $d$ is duration at rated power, $C_p$ is installed power cost, $C_{E,th}$ is theoretical installed energy cost (without factoring in efficiency losses and depth of discharge limitations), $\eta_D$ is discharge efficiency, $P_c$ is charging electricity price per cycle, $\eta_{RTE}$ is round-trip efficiency, VOM is variable operating and maintenance cost, FOM is fixed operating and maintenance cost, $C_{Re}$ is replacement cost of the energy storage medium, and L is cycle life of the storage system.

**Quantification and statistical analysis**

Statistical tools: parameter fitting was performed using Matlab linear regression.
Monte carlo simulation: random samples were generated to produce parameter distributions.
Uncertainty metrics: percentiles (e.g., 5th–95th) of ER were reported.

**Additional resources**

No clinical trial registrations or external resources were generated. The monte carlo simulation code developed in this study is available upon reasonable request.

**One-sentence descriptive title:** Table S1. Approaches to data collection and clarification of data sources in CAES projects.

***Excel Table S1 title:*** *Table S1. Approaches to data collection and clarification of data sources in CAES projects. In A-CAES systems, the overall system efficiency is typically defined as round-trip efficiency, which*

*is the ratio of energy output during the discharging phase to energy input during charging. The compressor and turbine performance are evaluated using isentropic efficiency, which indicates the deviation of actual thermodynamic processes from the ideal isentropic behaviours. For thermal energy storage units, thermal storage retention efficiency is used to quantify the proportion of stored thermal energy that remains usable after a given period, considering thermal losses. In contrast, D-CAES systems often employ cycle efficiency, defined as the ratio of useful work output over a complete thermodynamic cycle to the total energy input required for that cycle. Note that, the 2013 project is classified as A-CAES due to its adiabatic compression/expansion in the system [50]. Additionally, even for completed projects, costs can remain uncertain without detailed disclosure. For example, the Hydrostor who builds the Goderich plant claims a cost of $1000–3000/kW to build A-CAES plants with limited transparency. This cost range may present impact of location- and project-based costs across different A-CAES projects. Because the table is too extensive to be accommodated within the PDF, we have provided it as a standalone Excel file.*